\documentclass[fleqn,twoside]{article}
\usepackage{amssymb}
\usepackage{espcrc2}
\usepackage{graphicx}
\usepackage[figuresright]{rotating}

\newcommand{\AmS}{{\protect\the\textfont2
  A\kern-.1667em\lower.5ex\hbox{M}\kern-.125emS}}

\title{Measuring interface tensions in ${\rm 4d}$ ${\rm SU(N)}$ lattice
gauge theories}

\author{ Ph. de Forcrand
                \address[ETH]{Institute for Theoretical Physics, ETH Z\"urich,
                                CH-8093 Z\"urich, Switzerland}
                \address[CERN]{CERN, Physics Department, TH Unit, CH-1211 Gen\`{e}ve 23, Switzerland},
         B. Lucini
                \addressmark[ETH]
         and M. Vettorazzo
                \addressmark[ETH]
                \thanks{Talk presented by M. Vettorazzo}
        }

\begin{document}

\begin{abstract}
We propose a new algorithm to compute the order-order interface tension in ${\rm SU(N)}$
lattice gauge theories. The algorithm is trivially generalizable to a variety of models,
e.g., spin models. In the case $N=3$, via the \emph{perfect wetting} hypothesis, we can
estimate the order-disorder interface tension. In the case $N=4$, we study the ratio of
dual $k-$tensions and find that it satisfies Casimir scaling down to $T=1.2~T_c$.
\end{abstract}

\maketitle

\section{Introduction}

The phenomenon of phase-coexistence is the typical signature of a first-order transition.
The free energy of a system at the coexistence point is the sum of two contributions: the
bulk free energy $F_{\rm bulk}$, scaling like the volume $V$, and the free energy
$F_{\Sigma}$ of the \emph{interface} $\Sigma$ separating the two bulk phases. $F_{\Sigma}$
scales like an area. The interface tension $\sigma$ is defined as $\sigma=\lim_{\Sigma \to
\infty}\frac{F(\Sigma)}{\mid\Sigma\mid}$ (the \emph{reduced} tension is defined as
$\sigma^R=\frac{\sigma}{T}$). In the context of $SU(N)$ pure gauge theories ($N$ is the
number of colors) phase-coexistence occurs in two different situations: one at $T \geq T_c$
between ordered phases pointing in different directions in color space, the other at $T_c$,
between confined (`disordered') and deconfined (`ordered') phases. The corresponding
interface tensions are indicated $\sigma_{\rm oo}$ and $\sigma_{\rm od}$ respectively.

This paper presents a new algorithm to measure the order-order interface tension.
Nevertheless, we can also give an estimate of the order-disorder tension via the so-called
\emph{perfect wetting} hypothesis \cite{Frei:1989es}, which states that between two ordered
phases a layer of disordered phase can be generated at no free energy cost. Therefore the
free e\-ner\-gy of one order-order interface is related to that of \emph{two}
order-disorder interfaces, namely $F_{\rm oo}= 2F_{\rm od}$. In general, anyway,
$\sigma_{\rm oo}= w \sigma_{\rm od}, \hspace{0.1cm} w\leq 2$, because otherwise order-order
interfaces would be unstable. Therefore the choice $w=2$ gives a lower bound for
$\sigma_{od}$.

Our system is a $4d$ $SU(N)$ pure gauge theory at finite temperature with spatial periodic
boundary conditions (b.c.). $a$ indicates the lattice spacing. The volume is $V=L^3\cdot
L_t$; the Wilson action is used. We will indicate the elements of the center of $SU(N)$ by
$\zeta_i, \hspace{0.1cm}i=1,\ldots,N$. Heat-bath and over-relaxation are applied to $SU(2)$
subgroups of the $SU(N)$ matrices, according to the strategy used in \cite{Cabibbo:1982zn}.

\section{The method}

\begin{figure}[t]
\begin{center}
\includegraphics[height=3.5cm,angle=0.]{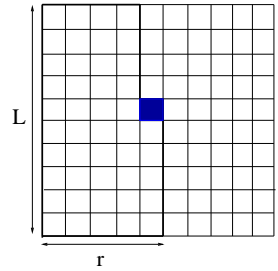}
\end{center}
\vspace{-1.0cm}\caption{\label{fig:central_plaq} Cartoon of the interface, in an
intermediate situation in which only half of the pla\-quet\-tes has been flipped. The solid
central plaquette is affected by the least finite size effects.} \vspace{-0.5cm}
\end{figure}

\begin{figure}[t]
\vspace{-0.5cm}
\begin{center}
\includegraphics[height=8.cm,angle=-90.]{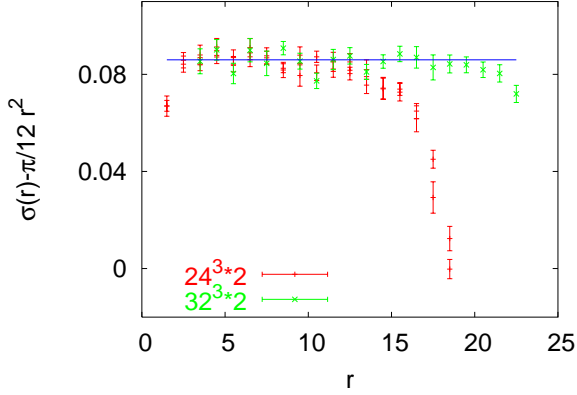}
\end{center}
\vspace{-1.0cm}\caption{\label{fig:plateau_SU3} (Minus log of the) Ratios in
Eq.(\ref{eq:chain_rule}) as a function of $r$, after removal of the L\"uscher correction. A
broad plateau develops from $r\sim \xi$ ($\xi=\frac{1}{\sqrt{\sigma}}\simeq 3.5$ is the
correlation length) to $r \to L$.}
\end{figure}
\begin{figure}[t]
\vspace{-0.5cm}
\begin{center}
\includegraphics[height=8.cm,angle=-90.]{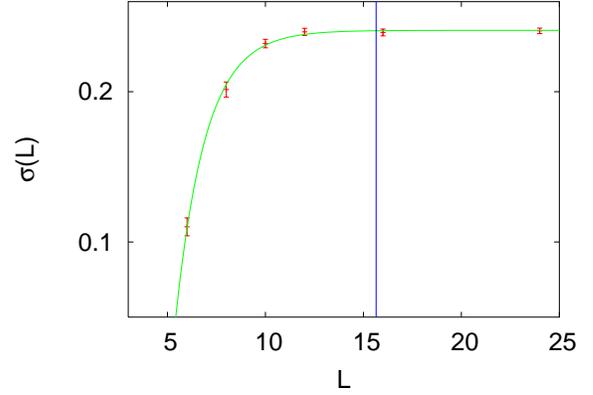}
\end{center}
\vspace{-1.0cm}\caption{\label{fig:FS_effects} Finite size effects on the interface tension
($L_t=2$, $\beta=5.15$). After removal of the L\"uscher correction, $\sigma$ is
systematically underestimated, unless lattices larger than $L \gtrsim 7/\sqrt{\sigma}$ are
used (vertical line). The curve is meant to guide the eye.}
\end{figure}

The algorithm we propose is an improvement of the so-called \emph{snake algorithm}
\cite{deForcrand:2000fi}. The `snake' idea is to add `by hand' a $2d$ interface in the
system by progressively flipping the coupling of a set of plaquettes dual to a  surface
$A$, according to the identity

\begin{equation}
\label{eq:chain_rule}
\frac{Z(A)}{Z(0)}=\frac{Z(A)}{Z(A-1)}\frac{Z(A-1)}{Z(A-2)}\ldots\frac{Z(1)}{Z(0)}
\end{equation}

\noindent where $Z(k), \hspace{0.1cm} k \in [0,\ldots A]$ indicates the partition function
of a system in which only $k$ plaquettes are flipped. The free energy of the interface is
$F(A)=\sigma^R_{oo}A=-\log\frac{Z(A)}{Z(0)}$. A direct measurement of $\frac{Z(A)}{Z(0)}$
is not possible due to a serious \emph{overlap problem}, which is alleviated by the
factorization Eq.(\ref{eq:chain_rule}). The price to pay is a very large number ($L^2$) of
independent simulations. Progress in this direction is made comparing the following
equations

\begin{equation}
\label{eq:decompose_the_area} e^{-\sigma A}=\prod_{A} e^{-\sigma a^2}, \qquad
\frac{Z(A)}{Z(0)}=\prod_{A} \frac{Z(k)}{Z(k-1)}
\end{equation}

\noindent the first just being a decomposition of the interface into the constituent
plaquettes, the second indicating that each ratio, up to finite size (F.S.) corrections,
contains all the information we need, namely

\begin{equation}
\label{eq:general_form_for_corrections} \frac{Z(k)}{Z(k-1)}=e^{-\sigma a^2}+( k-{\rm
dependent \hspace{0.1cm} F.S. \hspace{0.1cm} corrections})
\end{equation}

\noindent so that a \emph{single} simulation suffices. The gain in efficiency is ${\cal
O}(L^4)$: a factor $L^2$ in the number of simulations, another factor $L^2$ because the
$L^2$ variances of the ratios add up in Eq.(\ref{eq:chain_rule}). The observable is
measured in the following way:

\begin{equation}
\label{eq:measure_one_ratio} \frac{Z(k)}{Z(k-1)}=\frac{\langle e^{\beta \frac{1}{3}{\rm
Tr}(\zeta_1 \Pi_{k})}\rangle_k}{\langle e^{\beta \frac{1}{3}{\rm Tr}(\Pi_{k})}\rangle_k}
\end{equation}

\noindent where $\Pi_k$ indicates the $k-$th flipped plaquette, and the average $\langle
\cdot \rangle_k$ refers to the ensemble in which the first $(k-1)$ plaquettes are flipped,
the $k-$th plaquette has coupling zero, and all the others are unchanged. Further variance
reduction methods are described in \cite{deForcrand:2000fi}.

The leading finite-size corrections in Eq.(\ref{eq:general_form_for_corrections}) come from
Gaussian fluctuations of the interface \cite{Luscher:1980ac}. Our $r\cdot L$ interface
(Fig.~\ref{fig:central_plaq}) is periodic in one direction ($L$), and pinned in the other
($r$). The corresponding correction, given in terms of the Dedekind $\eta$-function,
reduces for $r \ll L$ to the Luscher-like $\sigma_{\rm eff}(r) \equiv -a^{-2} \log
Z(k)/Z(k-1) \approx \sigma + \frac{\pi}{12 r^2}$ \cite{Dietz:1982uc}. In
Fig.~\ref{fig:plateau_SU3} we show our measurements, after removal of this known
correction, as a function of $r$. A broad plateau develops, from small values $r \sim \xi$,
where $\xi=1/\sqrt{\sigma}$ is the correlation length, to large ones $r \to L$, showing
that additional corrections are very small. At very large distances, a systematic drop is
visible, because it becomes more favorable to produce a full, translationally invariant
interface plus a partial one of width $(L-r)$. Let us then fix $r=(\frac{L}{2}-1)$; after
removal of the L\"uscher corrections, Fig.~\ref{fig:FS_effects} shows that the tension
decreases considerably with the lattice size and reaches a plateau only when the empirical
condition $L
 \sqrt{\sigma} \gtrsim 7$ is satisfied.

\section{Results}

\begin{figure}[t]
\vspace{-0.5cm}
\begin{center}
\includegraphics[height=6.cm,angle=0.]{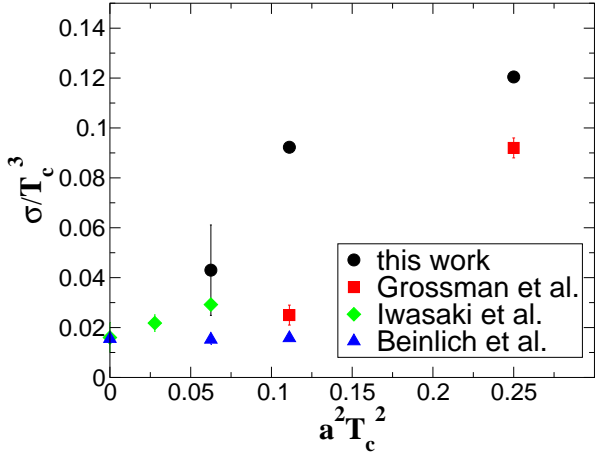}
\end{center}
\vspace{-1.0cm}\caption{\label{fig:tension_SU3} Summary of results on the order-disorder
interface tension for $SU(3)$. To avoid an underestimate of $\sigma$ because of F.S.
effects, we use a lattice of size $32^3\cdot 2$ ($a^2T_c^2=1/4$), $48^3\cdot 3$ ($a^2
T_c^2=1/9$), $64^3\cdot 4$ $(a^2T_c^2=1/16)$. The latter simulation is in progress.}
\end{figure}

In Fig.~\ref{fig:tension_SU3} we present preliminary results for the order-disorder
interface tension in $SU(3)$, together with a compilation of the published data. While our
$L_t=4$ simulation needs more statistics, our $L_t=2$ and $3$ determinations of $\sigma$
are accurate, and much larger than previous measurements obtained with the histogram method
\cite{Iwasaki:1993qu}\cite{Grossmann:1992dy}\cite{Beinlich:1996xg}\cite{Papa:1997ed}. We
assign this discrepancy to the smaller lattice sizes considered previously, which lead to a
systematic underestimate of $\sigma$ as in Fig.~\ref{fig:FS_effects}. A discussion of the
continuum limit is awaiting completion of the $64^3\cdot 4$ simulations.

\begin{table}[t]
\vspace{-0.cm}
\begin{center}
\begin{tabular}{|c|c|c|}\hline

$T/T_c$  & $L=16$ & $L=24$  \\

\hline

2.3 & 1.350(20)  & 1.342(13) \\ \hline

1.5 &     -      & 1.300(18) \\ \hline

1.2 & 1.277(33)  & 1.310(30) \\ \hline

\hline
\end{tabular}
\caption{\label{tab:SU4_preliminary_results} Ratio $\sigma_2/\sigma_1$ of $SU(4)$ interface
tensions ($L_t=5$). The Casimir perturbative value is $4/3$. For the data at $T=1.2~T_c$ we
find again that the F.S. effects are smaller than the statistical error only if $L \gtrsim
7/\sqrt{\sigma}$.}
\end{center}
\end{table}

In Tab.\ref{tab:SU4_preliminary_results} we present preliminary results for the case
$SU(4)$. In $SU(N)$ with $N>3$ we have more order-order tensions ($\sigma_k,
\hspace{0.05cm} k=1,\ldots, [N/2]$). In the weak coupling regime one can show the Casimir
relation $\frac{\sigma_k}{\sigma_1}=\frac{k(N - k)}{N - 1}$ \cite{Giovannangeli:2001b}.
This perturbative prediction seems accurate down to temperature $T=1.2~T_c$. Measurements
closer to $T_c$ are in progress.

\end{document}